\documentclass[preprint,12pt]{elsarticle}

\usepackage{setspace}
\usepackage[utf8]{inputenc} 

\usepackage[table]{xcolor}
\usepackage{subcaption}
\usepackage{balance}
\usepackage{amsmath,amssymb}
\usepackage{url}

\usepackage{comment}

\usepackage{color}
\usepackage{xcolor}

\usepackage{xspace}
\usepackage{graphicx}
\usepackage{longtable}
\usepackage{caption}
\usepackage{amsmath,amsthm}
\usepackage{multirow}
\usepackage{multicol}
\usepackage{booktabs}
\usepackage{url}
\usepackage{pifont}
\usepackage{color}
\usepackage{supertabular}
\usepackage[export]{adjustbox}

\usepackage{algorithm}
\usepackage{algorithmic}

\usepackage{amsmath}
\usepackage{amsthm,amsmath,amssymb}
\usepackage{mathrsfs}
\usepackage{bm}
\usepackage{subcaption}
\usepackage{hyperref}
\usepackage{CJKutf8}

\usepackage{cleveref}

\usepackage{cleveref}

\floatname{algorithm}{Algorithm}

\newcommand{\vct}[1]{\ensuremath{\boldsymbol{#1}}}
\newcommand{\mat}[1]{\ensuremath{\mathbf{#1}}}

\newcommand{\argmin}{\operatornamewithlimits{\arg\,\min}}

\newcommand{\myparagraph}[1]{\smallskip \noindent \textbf{#1}}
\newcommand{\ie}{{i.e.}\xspace}
\newcommand{\eg}{{e.g.}\xspace}
\newcommand{\etal}{{et al.}\xspace}

\newcommand{\imagenet}{{ImageNet}\xspace}
\newcommand{\dr}{{Diabetic Retinopathy Detection}\xspace}
\newcommand{\md}{{HAM10000}\xspace}

\begin{document}

\begin{frontmatter}

\title{Stateful Detection of Adversarial Reprogramming}

\author[NPU]{Yang Zheng}

\author[NPU]{Xiaoyi Feng}

\author[NPU]{Zhaoqiang Xia}

\author[NPU]{Xiaoyue Jiang}

\author[unica]{Maura Pintor}

\author[unica]{Ambra Demontis\corref{mycorrespondingauthor}}
\cortext[mycorrespondingauthor]{Corresponding author}
\ead{ambra.demontis@unica.it}

\author[unica]{Battista Biggio}

\author[NPU,unige]{Fabio Roli}

\address[NPU]{Northwestern Polytechnical University, Xi’an, China}
\address[unica]{University of Cagliari, Italy}
\address[unige]{University of Genoa, Italy}


\begin{abstract}

Adversarial reprogramming allows stealing computational resources by repurposing machine learning models to perform a different task chosen by the attacker. For example, a model trained to recognize images of animals can be reprogrammed to recognize medical images by embedding an adversarial program in the images provided as inputs. 
This attack can be perpetrated even if the target model is a black box, supposed that the machine-learning model is provided as a service and the attacker can query the model and collect its outputs. So far, no defense has been demonstrated effective in this scenario. 
We show for the first time that this attack is detectable using stateful defenses, which store the queries made to the classifier and detect the abnormal cases in which they are similar. Once a malicious query is detected, the account of the user who made it can be blocked. Thus, the attacker must create many accounts to perpetrate the attack. To decrease this number, the attacker could create the adversarial program against a surrogate classifier and then fine-tune it by making few queries to the target model. In this scenario, the effectiveness of the stateful defense is reduced, but we show that it is still effective. 


\end{abstract}

\end{frontmatter}



\section{Introduction}
\label{introduction}
Adversarial reprogramming is an attack that allows stealing the computational resources of machine learning models provided as a service by repurposing them to perform a task chosen by the attacker. For instance, an online service that uses a deep network to classify images of objects can be reprogrammed by attackers to solve CAPTCHAs\footnote{Completely Automated Public Turing test to tell Computers and Humans Apart.} automating the creation of SPAM accounts~\cite{elsayed19-ICLR}. Let us consider as an example the case depicted in Fig. \ref{fig:repr-success-and-failure}; where an attacker would like to repurpose a model trained to classify samples belonging to a \textit{source} domain (\eg, \imagenet objects) to classify samples belonging to a different, \textit{target} domain (\eg, the medical images of the HAM10000 dataset). To this end, the attacker should first establish a mapping function between the class labels of the source domain and those of the target domain (\eg, the first six classes of the ImageNet datasets could be associated to the class ``akiec'' of intraepithelial carcinoma, etc.). Once such a class mapping is established, the target-domain samples will be modified to embed the \textit{adversarial program}. Namely, a universal (equal for all the target-domain samples) adversarial perturbation, optimized to have such samples assigned to the desired source-domain classes. For a more detailed explanation of how this attack works, we refer the reader to Sec. \ref{sec:adversarial-reprogramming}. 
The first work that proposed adversarial reprogramming~\cite{elsayed19-ICLR} assumed that the attacker knows the architecture and weights of the target model (the so called white-box scenario), which is seldom true. However, recent results~\cite{tsai20-pmlr} showed that adversarial reprogramming can be executed even if the attacker has only query access to the target model (black-box scenario). Namely, if the attacker can only make a sequence of queries to the target model and collect the outputs. So far, no defense has shown to be effective in this scenario; therefore, it remains unclear if and to which extent this attack can be mitigated.

\begin{figure}[!t]
	\centering
    \includegraphics[width=0.60\textwidth]{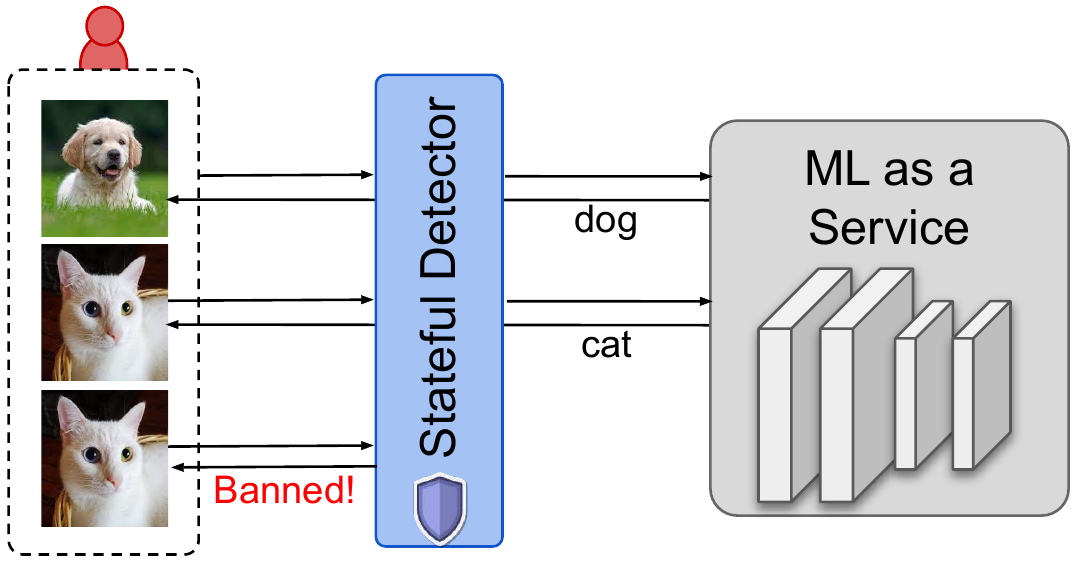}
	\caption{A machine learning (ML) system provided as a service and protected by a stateful detector which analyzes the queries made by a user's account and eventually bans it.}
	\label{fig:fig1}
\end{figure}

In this work, we show for the first time that this attack can be effectively mitigated with stateful defenses (Sec. \ref{sec:stateful-defenses}). In a black-box scenario, the attacker can estimate the gradient needed to optimize the adversarial program with numerical techniques. To this end, she has to make numerous queries with inputs quite similar to each other. This similarity can be exploited to detect reprogramming attacks using a stateful detector. This defense was originally proposed by Chen et al. ~\cite{chen2020stateful} to protect machine learning systems against a different kind of attack, namely the evasion attack: an attack that computes a perturbation that, if applied to a single sample, allows the attacker to have it classified as the desired class. 
Stateful detectors record the queries made by a user to the classifier and store them in a temporary history buffer. For each new query, if the detector finds many old queries quite nearby, it will flag the new query as an attack, and the system will block that user (see Figure \ref{fig:fig1}). The attacker will thus has to create another account to be able to continue optimizing the adversarial perturbation. Therefore, this defense substantially increases the effort that the attacker should make. 

Our experimental analysis (Sec. \ref{sec:experimental-analysis}) shows that stateful defenses are highly effective against black-box reprogramming. However, it is worth noting that that the attacker could reduce the number of queries by leveraging a known property of attacks called (\textit{transferability}), namely, the capability of an attack computed against a given model (\textit{surrogate}) to be effective against a different (\textit{target}) model~\cite{demontis19-usenix}. In our experiments, we have tested the effectiveness of adversarial reprogramming when the attacker tries to leverage this property by computing the adversarial program against a surrogate model and then fine-tuning it by querying the target model. Our results show that the attacker can reduce the number of queries and increase the success of the attack by exploiting  (\textit{transferability}); however, our defense remains effective and is a valuable deterrence mechanism. 

We conclude this paper by discussing related work (Sec.~\ref{sec:related_work}), our main contributions, the limitations of our work, and promising directions for future work (Sec.~\ref{sec:conclusion}).

\begin{figure}[!t]
	\centering
	\includegraphics[width=0.999\textwidth]{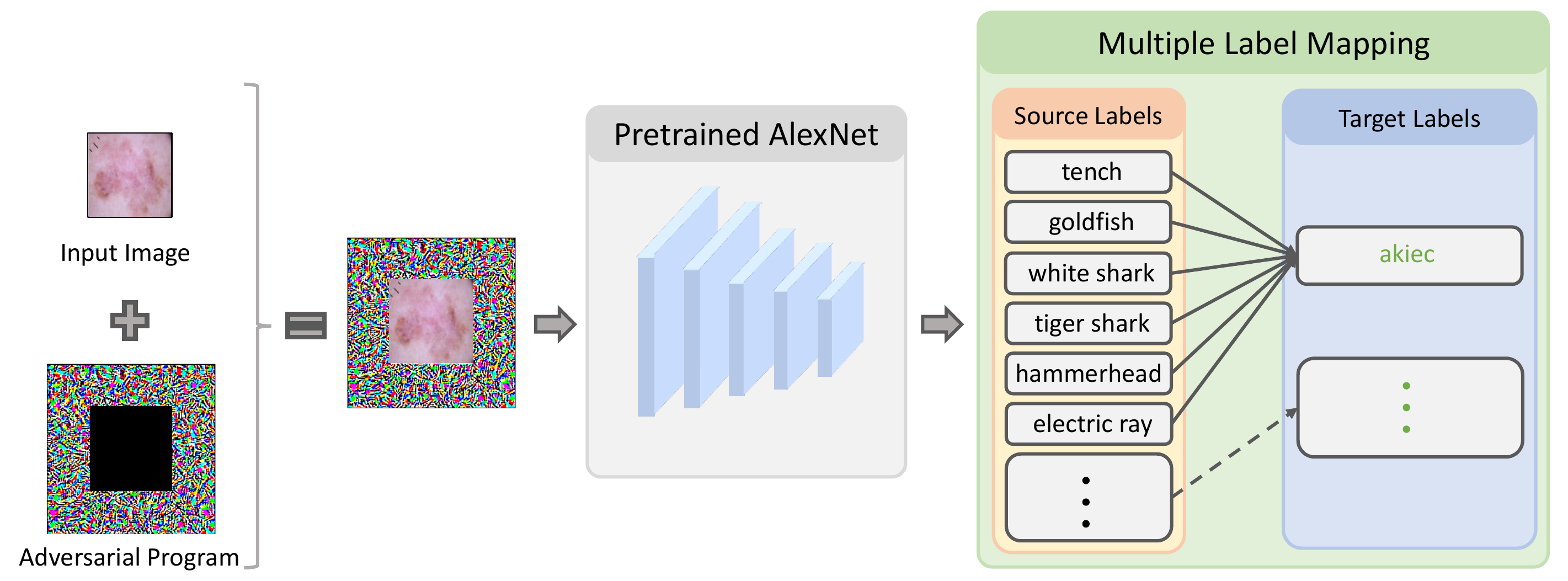}
	\caption{Adversarial reprogramming of AlexNet~\cite{alex12-nips}, trained on the \imagenet dataset. Multiple source-domain classes are mapped to one of target-domain classes (\eg, \textit{``tench'', ``goldfish'', ``white shark'', ``tiger shark'', ``hammerhead'', ``electric ray''} are mapped in \textit{``akiec''}, etc.). In the example, the \textit{\md} image ``\textit{akiec}'' is embedded in the adversarial program, and classified by pretrained AlexNet as desired, namely as one of the source-domain classes that is mapped to the ``\textit{akiec}'' class.}
	\label{fig:repr-success-and-failure}
\end{figure}

\section{Adversarial Reprogramming}
\label{sec:adversarial-reprogramming}
In this section, we first explain that the task of developing an adversarial program can be mathematically formulated as an optimization problem. Then, we describe the algorithm that the attacker can use to solve this problem when the target model is a black box, that is the scenario considered in this work (Sec. \ref{sec:black-box-programs}).

\subsection{Problem Formulation}
In this section we introduce the mathematical model of adversarial reprogramming. Let us assume that we have a source-domain dataset $\mathcal{S}=(\tilde{\vct x}_j, \tilde{y}_j)_{j=1}^m$ and a target-domain dataset $\mathcal{T} = ( \vct x_i, y_i)_{i=1}^n$, consisting of $m$ and $n$ samples, along with their labels. 
The samples of the source and target domain are represented as vectors, respectively in $\mathcal{\tilde{X}} = [-1,1]^{d \times d \times 3}$ and $\mathcal{X} = [-1,1]^{d^\prime \times d^\prime \times 3}$. 
The class labels belong to different domains, respectively, $\tilde{y} \in \mathcal{{\tilde Y}}$ for the source domain, and $y \in \mathcal{Y}$ for the target domain. 
Let $s$ be the total number of classes in the source-domain dataset and $\mathcal{{\tilde Y}} = \{ \tilde{y}_1,\ldots,\tilde{y}_k,\ldots,\tilde{y}_s \}$ be the set of the source labels. We define the target model that we would like to reprogram as $f : \tilde{\mathcal{X}} \mapsto \mathbb{R}^s$. This model is parameterized by $\vct \theta \in \mathbb{R}^t$ and provides as output a vector of confidence scores $f(\tilde{\vct{x}}_j, \vct \theta) = \{c_{\tilde{y}_1}(\tilde{\vct{x}}_j, \vct \theta),\ldots,c_{\tilde{y}_k}(\tilde{\vct{x}}_j, \vct \theta),\ldots,c_{\tilde{y}_s}(\tilde{\vct{x}}_j, \vct \theta)\}$. 
To reprogram it we should define a mapping between the source- and the target- domain class labels, \eg using the Multiple Label Mapping (MLM) proposed by Tsai \etal~\cite{tsai20-pmlr}. Let $t$ be the total number classes of the target-domain dataset and $\mathcal{Y}= \{y_1,\ldots,y_i,\ldots,y_t\}$ be the set of the target domain labels. We can define a MLM function $h_{y_i \in \mathcal{Y}}(f(\tilde{\vct{x}}_j, \vct \theta)) = \frac{1}{\mid \vct K\mid} \sum_{k \in \vct K} c_k(\tilde{\vct{x}}_j, \vct \theta)$, where $ \vct K \subseteq{\tilde{\mathcal{Y}}}$ is the subset of source labels, and $\mid \vct K\mid$ is the number of elements of $ \vct K$, that maps a subset of multiple-source labels to a one-target label (\eg, the source-domain label set \{\textit{``tench'', ``goldfish'', ``white shark'', ``tiger shark'', ``hammerhead'', ``electric ray''}\} is mapped to the target-domain label \textit{``akiec''} as in Fig. \ref{fig:repr-success-and-failure}). 

\begin{figure*}[!t]
	\centering
	\includegraphics[width=0.80\textwidth]{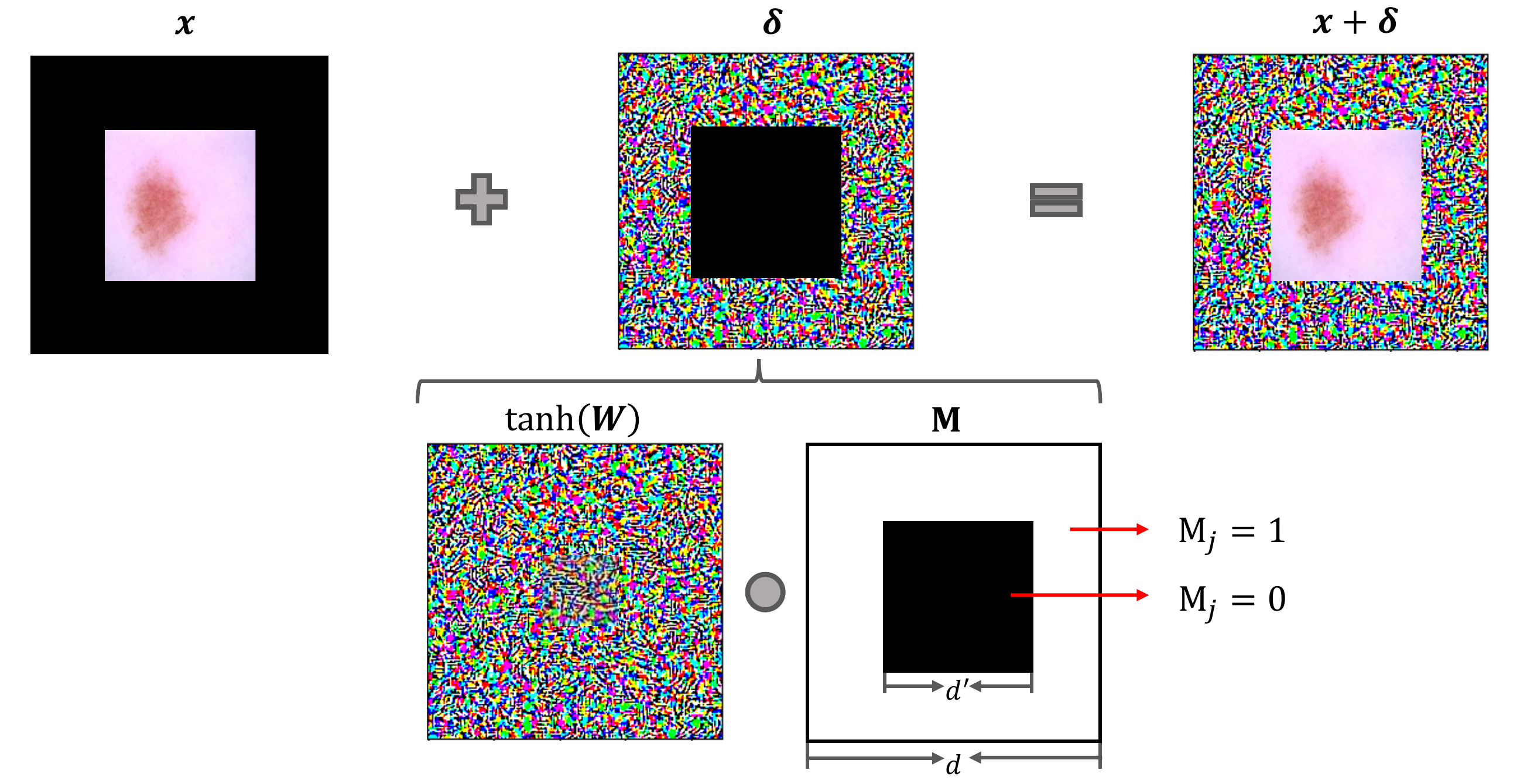}
	\caption{Reprogramming mask $\mat M$ used to restrict the adversarial perturbation ($\vct \delta$) to the frame surrounding the target-domain input image ($\vct x$), initially padded with zeros.}
	\label{fig:mask}
\end{figure*}

\myparagraph{Reprogramming Mask.} In this work, we focus on programs consisting of a frame surrounding the target-domain samples as shown in Fig.~\ref{fig:repr-success-and-failure}, also considered in the seminal work that proposed adversarial reprogramming~\cite{elsayed19-ICLR}. 
This means that the target-domain samples are assumed to be smaller than the source-domain samples, \ie $d^\prime < d$, and padded with zeros to reach the input size of the target model. For example, the images of the \md dataset consist of $200 \times 200 = 40,000$ pixels per channel, and should be padded with 10,176 zeros per channel to reach the input size of \imagenet models (which have $224 \times 224 = 50,176$ pixels per channel).
To compute the adversarial programs, we use a reprogramming mask (shown in Fig.~\ref{fig:mask}): a binary vector $\mat M \in \{0,1\}^{d}$ 
whose values are set to 0 in the region occupied by the target-domain samples
, and to 1 in the surrounding frame.

\myparagraph{Adversarial Program.} Under these assumptions, we can define the adversarial program $\vct \delta$ as:
\begin{equation}
	\label{eq:program}
	\vct \delta=\tanh(\vct W\circ \mat M)=\tanh(\vct W)\circ \mat M,
\end{equation}
where $\vct W\in \mathbb{R}^{d \times d \times 3} $ is a vector containing the adversarial program parameters to be learned, 
the $\circ$ operator denotes element-wise vector multiplication, and the ``$\tanh$'' function constrains the adversarial program in the feasible domain $\mathcal{X}=[-1,1]^{d \times d \times 3}$.

\myparagraph{Loss Function.} The optimal adversarial program $\vct \delta^\star$ can be obtained by solving the following optimization problem:
\begin{equation}
	\label{eq:reprogramming1}
	\vct \delta^\star \in \argmin L(\vct \delta, \mathcal{T}) = \frac{1}{n} \sum_{i=1}^n \ell(h_{y_i}(f(\vct x_i+ \vct \delta, \vct \theta)), y_i),
\end{equation}
where $\vct \delta$ is the optimized adversarial program, $h_{y_i}(\cdot)$ is the MLM function, $\ell$ is the focal loss \cite{Lin17-ICCV}, which takes on high positive values when the perturbed target-domain samples are not confidently assigned to the desired target-domain label. 

\subsection{Solution Algorithm}
Supposing to have complete access to the target system (white-box attack scenario),
the optimization problem in Eq.~\eqref{eq:reprogramming1} can be solved with the Algorithm~\ref{alg:reprogramming} which extends the Gradient Descent (GD) algorithm. 
This algorithm iteratively (line~\ref{line:forloop}) updates the adversarial program $\vct \delta$ to minimize the expected loss on the target-domain samples. In each iteration, the target-domain samples are randomly shuffled (line~\ref{line:shuffling}) and subdivided into $b$ batches. The adversarial program is then updated by iterating over the batches (line~\ref{line:forloopbatches}). To this end, first, the gradient $\vct g$ (line~\ref{line:avg_grad}) is computed by averaging the ones obtained considering each single sample. The gradient for the $i^{\rm th}$ sample is computed as $\vct g_i = \nabla_{\vct x} \ell(h_{y_i}(f(\vct x_i+ \vct \delta, \vct \theta)), y_i)$. Then, the adversarial program parameters are updated with an $\eta$-sized step (line~\ref{line:gradient_step}) in the the steepest descent direction (the opposite of $\vct g$). 
After updating $\vct \delta$, the algorithm constrains the program to be onto the feasible space $\mathcal{X} =[-1, 1]^{d\times d \times 3}$ employing the function $\tanh(\cdot)$ (line~\ref{line:proj_step}). 
The algorithm finally returns the adversarial program $\vct \delta^\star$ that achieves the minimum classification loss across the whole optimization process (line~\ref{line:return}).

\begin{algorithm}[t]
	\caption{Adversarial Reprogramming}
	\begin{algorithmic}[1]
		\REQUIRE the target-domain dataset $\mathcal{T}=(\bm{x}_i, {y}_i)^n_{i=1}$
		the model parameters $\vct \theta$, the batch size $B$, the adversarial program parameters $\vct W$, the reprogramming mask $\mat M$, the number of iterations $N$, the step size $\eta$, and the function $\tanh(\cdot)$.
		\ENSURE the optimal adversarial program $\vct \delta^*$.
		\STATE $\vct \delta \gets \vct 0$, $\text{loss}_{ \vct \delta*} \gets \infty$\label{line:init}, Randomly initialize $\vct W$ 
		\STATE $t \gets 0$
		\FOR {$t<N$}\label{line:forloop}
		\STATE \text{Randomly shuffle the samples in }$\mathcal{T}$\label{line:shuffling} 
		\STATE $b \gets 0$
		\FOR{$b < \lfloor{\frac{n}{B}} \rfloor$}\label{line:forloopbatches}
		\STATE $\vct g \gets \frac{1}{B} \sum_{i=B\cdot b}^{B\cdot b +B-1} \vct g_i$ \label{line:avg_grad}
		\STATE $\vct W \gets \vct W -\eta \times \vct g $ \label{line:gradient_step}
		\STATE $\vct \delta \gets \tanh(\vct W\circ \mat M)$ \label{line:proj_step}
		\STATE $b \gets b + 1$ 
		\ENDFOR
		\STATE $\text{loss}_{\vct \delta} = L(\vct \delta, \mathcal{T})$ \text{(compute loss as given in Eq.~\ref{eq:reprogramming1})} \label{line:loss}
		\IF{$\text{loss}_{\vct \delta} < \text{loss}_{ \vct \delta*}$}\label{line:check_stop}
		\STATE $\vct \delta^* \gets \vct \delta$\label{line:delta_update}
		\STATE $\text{loss}_{\vct \delta*} \gets \text{loss}_{\vct \delta}$
		\ENDIF
		\STATE $t \gets t + 1$ 
		\ENDFOR
		\STATE {\bfseries return} $\vct \delta^\star$ \label{line:return}
	\end{algorithmic}
	\label{alg:reprogramming}
\end{algorithm}

\subsection{Black-box Adversarial Reprogramming}
\label{sec:black-box-programs}
The methodology we explained in the previous section assumes that the attacker has full knowledge of the target system (i.e., it knows its architecture and weights) and thus can compute the gradient of the loss function w.r.t the input samples. However, this is hardly ever true because online machine-learning services avoid disclosing information about their machine learning algorithms. Often the attackers have no information about the target system (back-box scenario). They know only the task (\ie, image classification, object detection, malware classification, etc.) and have an idea of which potential transformations they can apply to the input to cause some feature changes~\cite{tenyears}. For example, the attackers know that its input features represent image pixels; thus, the input features can assume any value suitable for images' pixels.

In the black-box scenario, the gradients needed in line~\ref{line:avg_grad} of Algorithm \ref{alg:reprogramming} cannot be analytically-computed by the attacker. Nevertheless, the attacker can still execute reprogramming attacks~\cite{tsai20-pmlr}. Querying the target system and collecting outputs (\ie, provided labels, confidence scores), the attacker can estimate the required gradient with numerical techniques. 
Using one-sided averaged gradient estimators \cite{liu_18-nips, tu2019autozoom}, as done in \cite{tsai20-pmlr}, the gradient for the $i^{\rm th}$ sample can be estimated as:

\begin{equation}
	\label{eq:estimate}
	\begin{aligned}
	\hat{\vct g}_i = \frac{b}{q\mu} \sum_{j=1}^q [\ell(h_{y_i}(f(\vct x_i+  \mu \vct u_{ij}, \vct \theta)), y_i) \\ - \ell(h_{y_i}(f(\vct x_i, \vct \theta)), y_i)] \vct u_{ij}, 
	\end{aligned}
\end{equation}

where $b$ is a tunable scaling parameter that balances the bias and variance trade-off of the gradient estimation error, $\mu > 0$ is a smoothing parameter, $q>0$ is the parameter that influences the number of queries, and $\{ \vct u_{ij}\}_{j=1}^q \in \mathbb{R}^{d\times d \times 3}$ are i.i.d. random directions drawn from a uniform distribution over a unit sphere.

\section{Stateful Defenses}
\label{sec:stateful-defenses}
This section explains the working principles behind stateful defenses and how an attacker could reduce their effectiveness. 

\subsection{Query Detection}
As we explained in the previous section, to perform adversarial reprogramming in a black-box scenario, the attacker has to send many queries with inputs that are quite similar to each other to the target model. This is required to estimate the gradients as explained in  Eq. \ref{eq:estimate}. Therefore, we conjecture that this attack can be easily detected with the stateful detector proposed by Chen et al.~\cite{chen2020stateful}. The key hypothesis of the Chen et al. method implies that the sequence of queries used to generate a black-box attack are distinguishable from the ones usually made by benign users. Based on this hypothesis, the authors proposed a defense that relies on the observation that existing black-box attacks often make a sequence of highly self-similar queries (\ie, each query in the sequence is highly similar to some prior queries in the sequence). 

\begin{figure*}[!t]
	\centering
	\includegraphics[width=0.90\textwidth]{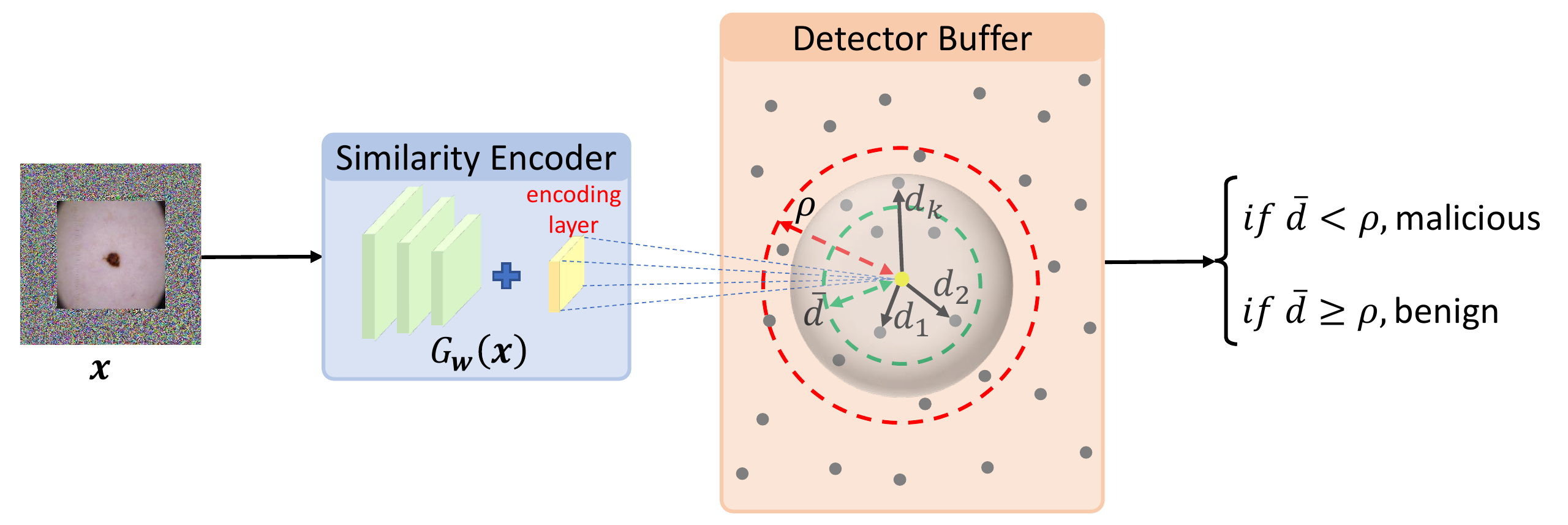}
	\caption{A high-level representation of a stateful defense. The detector maps the received queries in a low-dimensional space using a similarity encoder ($ G_{\vct w}(\vct x)$) and stores them into a temporary buffer. If the average distance ($\bar{d}$) between the current query (in yellow) and its $k$-nearest neighbor queries in the buffer (the points inside the gray sphere) is lower than a threshold ($\rho$), the query is flagged as malicious.}
	\label{fig:detector}
\end{figure*}

As shown in Fig.~\ref{fig:detector}, the detector, for each query ($\vct x$) received by a user: (i) Maps it into a low-dimensional space using a similarity encoder $G_{\vct w}(\vct x)$; (ii) Check how many queries the detector has already stored in its memory buffer ($Q$); (iii) If the detector has already memorized at least $k$ queries ($Q>k$), first, it computes the average $l_2$ distance in the low-dimensional space between the new query (the yellow point in Figure~\ref{fig:detector}) and the $k$-nearest queries memorized (the ones inside the gray sphere in Figure~\ref{fig:detector}); Then, if the computed average distance is smaller a chosen threshold $\rho$, the detector flags this sequence of queries as an attack. The user that made these queries will be blocked, and the attacker will have to create another account to issue more queries to refine the attack. To assess the detection capabilities in this scenario, for simplicity, we will consider all the queries made by the attacker as coming from the same user. However, to assess the detector performance fairly, we will reset the memory buffer whenever an attack is detected. In the real world, the attacker will have to change the account, and thus we will not know that the queries made by the two users, namely the new user and the old one, are actually made by the same person (the attacker). 

In the following, we propose a metric that can be used to assess the detectors' success. To this end, let us consider a detector that has just stored the first $k$ queries made by a user. When the user makes its $k+1$ query, if the average distance of this query from the $k$ stored queries falls below a chosen threshold, the detector will consider it an attack. Upon detection, the number of detected malicious queries will be equal to ``1'', and the buffer containing the previous $k+1$ queries will be cleared. Otherwise, the detector will focus on the next query. Therefore, the number of detections is $D \leq \lfloor \frac{Q}{k+1} \rfloor$, where the latter is the maximum number of possible detections, given that the detector should collect at least $k$ queries before examinating them to detect attacks. To clarify this point, we explain in details how the detector will work on the following ten queries ($Q=10$): $(\vct a, \vct b, \vct c, \vct d, \vct e, \vct f, \vct g, \vct h, \vct i, \vct j)$, when the number of detections $D=0$, and the parameter $k=3$. The detector will start checking for attacks once it have stored at least $k$ queries in its buffer. In the presented case, the queries $(\vct a, \vct b, \vct c)$. Then, once it will have receive the $k+1$ query ($\vct d$), the detector will compute the average distance between $\vct d$ and $\vct a$, $\vct b$, $\vct c$. Let $\mathcal{D}(\vct i, \vct j)$ represent the distance of $\vct i$ and $\vct j$. The detector will compute $\mathcal{D}(\vct a, \vct d)$, $\mathcal{D}(\vct b, \vct d)$, $\mathcal{D}(\vct c, \vct d)$, and the average distance $\bar{\mathcal{D}}_{\vct{abcd}} =\frac{\mathcal{D}(\vct a, \vct d) + \mathcal{D}(\vct b, \vct d) + \mathcal{D}(\vct c, \vct d)}{3}$. If $\bar{\mathcal{D}}_{\vct{abcd}}$ is smaller than $\rho$, the detector will flag the queries $(\vct a, \vct b, \vct c, \vct d)$ as malicious. Thus will add ``1'' to the number of the detections, \ie $D=1$, and then will clear its memory buffer. Otherwise, the detector will continue to compute the distance of the next new input $\vct e$ with $(\vct a, \vct b, \vct c, \vct d)$, \ie 
$\mathcal{D}(\vct a, \vct e)$
$\mathcal{D}(\vct b, \vct e)$, $\mathcal{D}(\vct c, \vct e)$, $\mathcal{D}(\vct d, \vct e)$. Then, if the average distance with the 3 nearest queries is smaller than $\rho$, the detector will flag the queries as an attack and add ``1'' to the number of the detections, \ie $D=1$, then will clear its buffer. It is not difficulty to see that the stateful defense is based on groups rather than individual queries, where the size of each group is $k+1$ (the maximum number of detection is one every $k+1$ samples), thus in our example $D \leq 2$. Therefore, to evaluate the performance of stateful defense, we introduce the detection coefficient $\sigma$ computed as follows:

\begin{equation}
	\label{eq:detecting_rate}
	\sigma = \frac{D}{Q} \leq \frac{1}{k+1}. 
\end{equation}

From the Eq.~\eqref{eq:detecting_rate}, we can see $\sigma$ is only related to $k$. Consequently, $k$ plays an important role in the scheme of stateful defense. To obtain a value $\in [0, 1]$, we define $\sigma ^{\star}$ as the normalization of $\sigma$:

\begin{equation}
	\label{eq:normalization_rate}
	\sigma ^{\star} = (k+1)\sigma. 
\end{equation}

From Eq.~\eqref{eq:normalization_rate}, it's not difficulty to see that $\sigma ^{\star} \in [0, 1]$, where $\sigma ^{\star}=0$ when $D=0$, and $\sigma ^{\star}=1$ when $D= \frac{Q}{k+1}$, which is the maximum number of detections that we can have for a chosen $k$. 

\subsection{Dimensionality Reduction.}
\label{dimreduction}
As we have explained before, to detect attacks, the detector evaluates the $l_2$-norm distance of the current query with the ones stored in its buffer. Their distance in input space would be computationally expensive to compute and would not be significant. E.g., a small rotation or translation can cause dramatic distance changes. Therefore, we consider their distance in a space with a smaller dimensionality. Following Chen et al.~\cite{chen2020stateful}, we use a similarity encoder to reduce the dimensionality. 

\begin{figure*}[!t]
	\centering
	\includegraphics[width=0.999\textwidth]{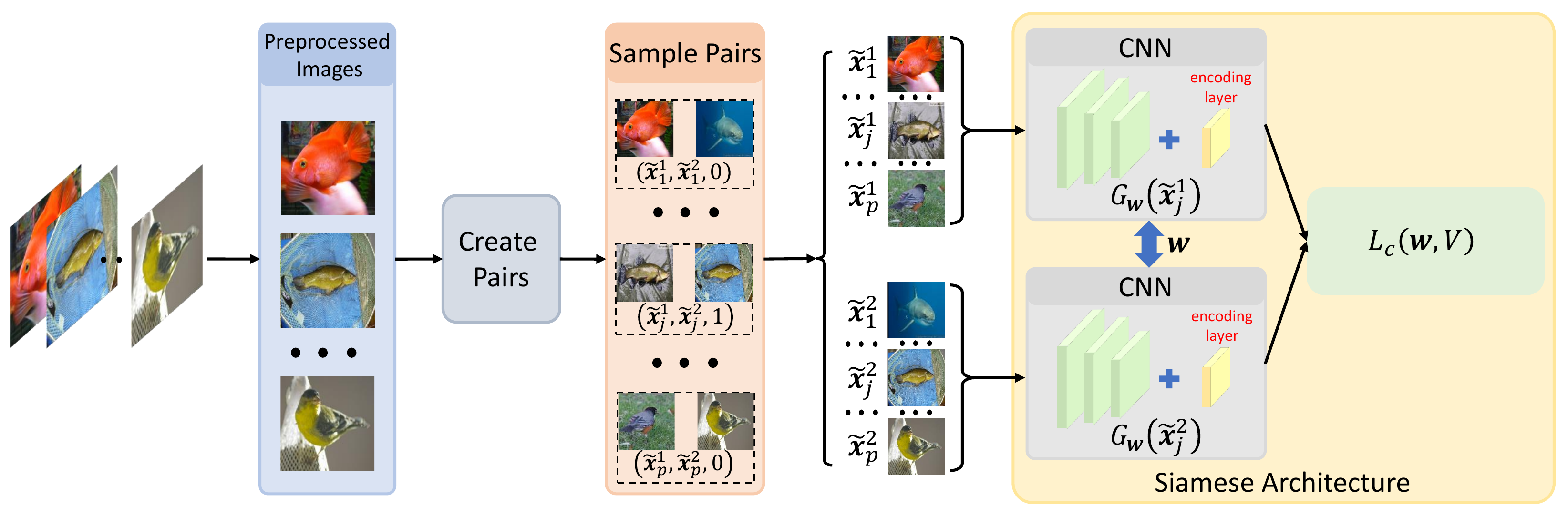}
	\caption{A high-level representation of the procedure used to train the similarity encoder. The samples of the source dataset are first paired. Then, the similarity encoder is trained on pair of samples to find a mapping in which the samples belonging to the same class are near, whereas the ones belonging to different classes are far from each other.}
	\label{fig:encoder}
\end{figure*}

To create the similarity encoder, as shown in Fig.~\ref{fig:encoder}, we use a \textit{Siamese architecture}. The Siamese architecture is constituted by two neural networks initialized with the same weights. We follow~\cite{hadsell2006dimensionality} and~\cite{chopra2005learning} to create the sample pairs (see Fig.~\ref{fig:encoder}) required to train it.
We construct a training dataset $V= (\vct v_j, l_j)_{j=1}^{p}$, where $\vct v_j = (\tilde{\vct x}_j^1 \in \mathbb{R}^{d\times d\times3},\tilde{\vct x}_j^2\in \mathbb{R}^{d\times d\times3})$ is the $j$-th sample pair made by samples of the source-domain dataset, $l_{j}$ is the label of the $j$-th sample pair, where $l_{j}=0$ ($l_{j}=1$) if $\tilde{\vct x}_j^1$ and $\tilde{\vct x}_j^2$ are similar, namely they belong to the same class (dissimilar, namely they belong to different classes), and $p$ is the number of possible pairs of samples belonging to the source domain dataset. Let $\vct w$ be the weights  shared by the two networks $G_{\vct w} (\cdot)$. 
As shown in Eq.~\eqref{eq:distance}, we compute the $l_2$ distance between $G_{\vct w} (\tilde{\vct x}_j^1)$ and $G_{\vct w} (\tilde{\vct x}_j^2)$, namely between two samples in the lower-dimensional space produced by the similarity encoder as:

\begin{equation}
	\label{eq:distance}
	DS_{\vct w}(\vct v_j) = {\Vert G_{\vct w} (\tilde{\vct x}_j^1) - G_{\vct w} (\tilde{\vct x}_j^2) \Vert}_2. 
\end{equation}

As in~\cite{hadsell2006dimensionality} we train the siamese architecture that we use to implement the similarity encoder with the contrastive loss function:

\begin{equation}
	\label{eq:ls_encoder}
	L_c(\vct w, V) = \sum_{j=1}^p \ell_c(\vct w, \vct v_j, l_j),
\end{equation}

\begin{equation}
	\begin{aligned}
	\label{eq:ls_encoder1}
		\ell_c(\vct w, \vct v_j, l_j) = & \underbrace{(1-l_j)\frac{1}{2}[DS_{\vct w}(\vct v_j)]^2}_{l_s}+ \\& \underbrace{l_j\frac{1}{2}\{\max[0, z-DS_{\vct w}(\vct v_j)]\}^2}_{l_d}, 
	\end{aligned}
\end{equation}
where $l_s$ is the partial loss function for similar pairs, $l_d$ is the partial loss function for dissimilar pairs, and $z$ is a margin. The $l_s$ term encourages the similarity encoder to find the weights for which similar pairs are mapped near each other. The $l_d$ term enables it to find weights for which the dissimilar pairs have a distance lower than the chosen margin $z$. 

\subsection{Leveraging Transferability to Defeat Stateful Defenses}
Stateful defenses allow detecting malicious queries and consequently blocking the attacker's account. To execute adversarial reprogramming, the attackers will have thus to create multiple accounts. To reduce the number of detections and thus accounts that they have to create, attackers might leverage on a property of attacks called transferability, namely the ability of an attack computed against a model (surrogate) to be effective against a different (target) model~\cite{demontis19-usenix}. An attacker can exploit this property to craft an adversarial program with the white-box reprogramming attack against a surrogate model, thus avoiding making any malicious query to the target model. However, if the surrogates and the target model are not similar, the computed adversarial program might not be sufficiently accurate in reprogramming the target model. Nevertheless, it might be exploited by the attacker. The attacker might use it to initialize the adversarial program and then fine-tune it by making few queries to the target model. In the next Section, we evaluate by experiments at which extent one attacker could exploit the transferability of attacks to defeat our stateful defence.

\section{Experimental Analysis}
\label{sec:experimental-analysis}
In the following, we describe the experimental setup used in the experiments we made to assess the effectiveness of stateful defenses, then we report the experimental results.

\subsection{Experimental Setup}
\label{experiment_setup}



In the following, we describe all the details required to replicate our experiments. 

\myparagraph{Datasets.} 
Because the \imagenet large-scale training corpus has gained popularity in computer vision as an evaluation benchmark and many pre-trained architectures are available on the Internet, we choose \imagenet as our source-domain dataset. Adversarial reprogramming has been recently shown in \cite{tsai20-pmlr} to be particularly useful for reprogramming models for datasets containing only few samples. This is usually true for medical datasets as their samples are quite costly to collect; thus, only a few samples are usually available. Therefore, we have chosen the two medical datasets used in that work as our target domain datasets. We present the details of the dataset used in the following.\newline

\begin{figure*}[!t]
	\centering
	\subfloat[]{\includegraphics[width=0.45\textwidth]{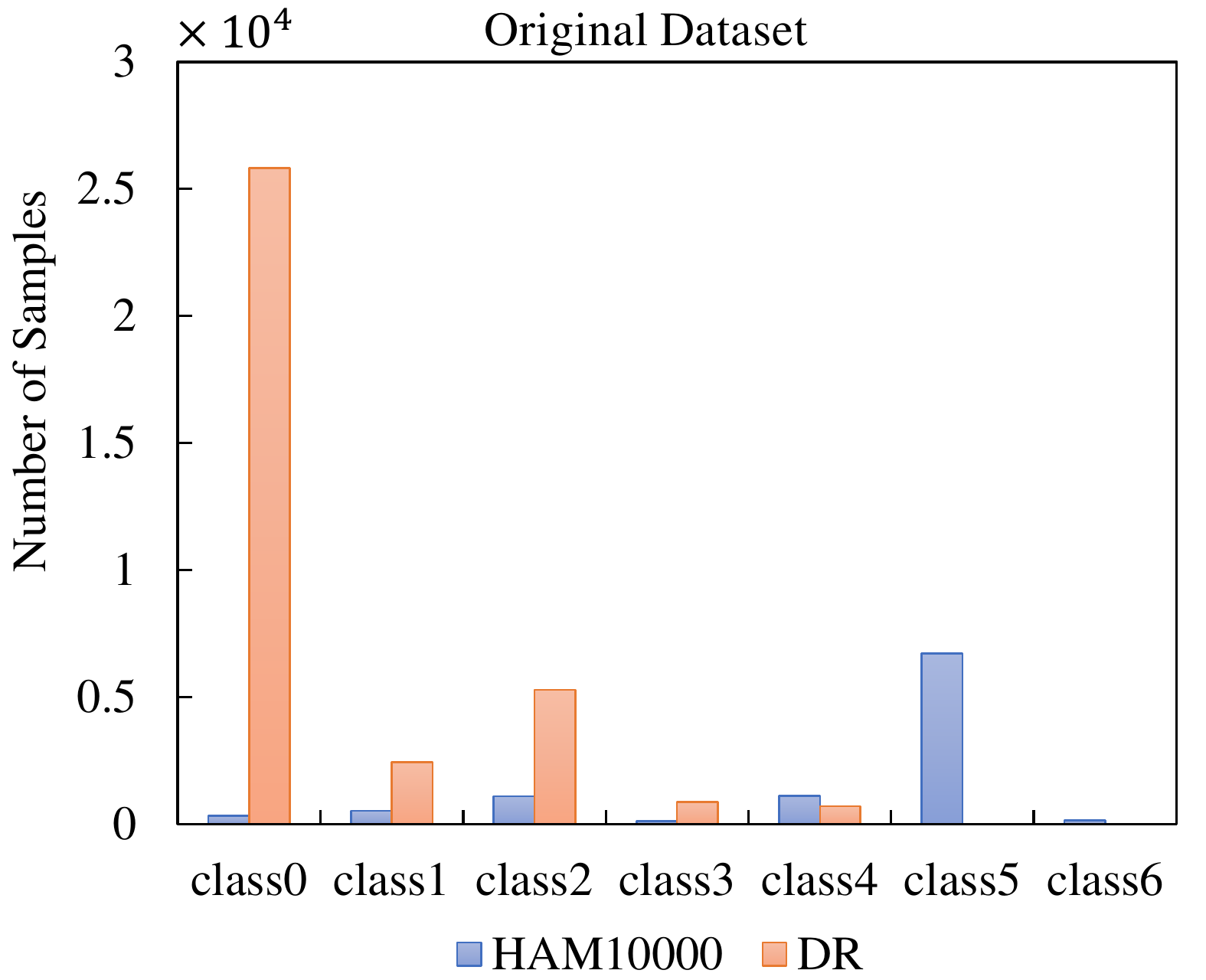}
	\label{fig:original_dataset}}
	\hfil
	\subfloat[]{\includegraphics[width=0.45\textwidth]{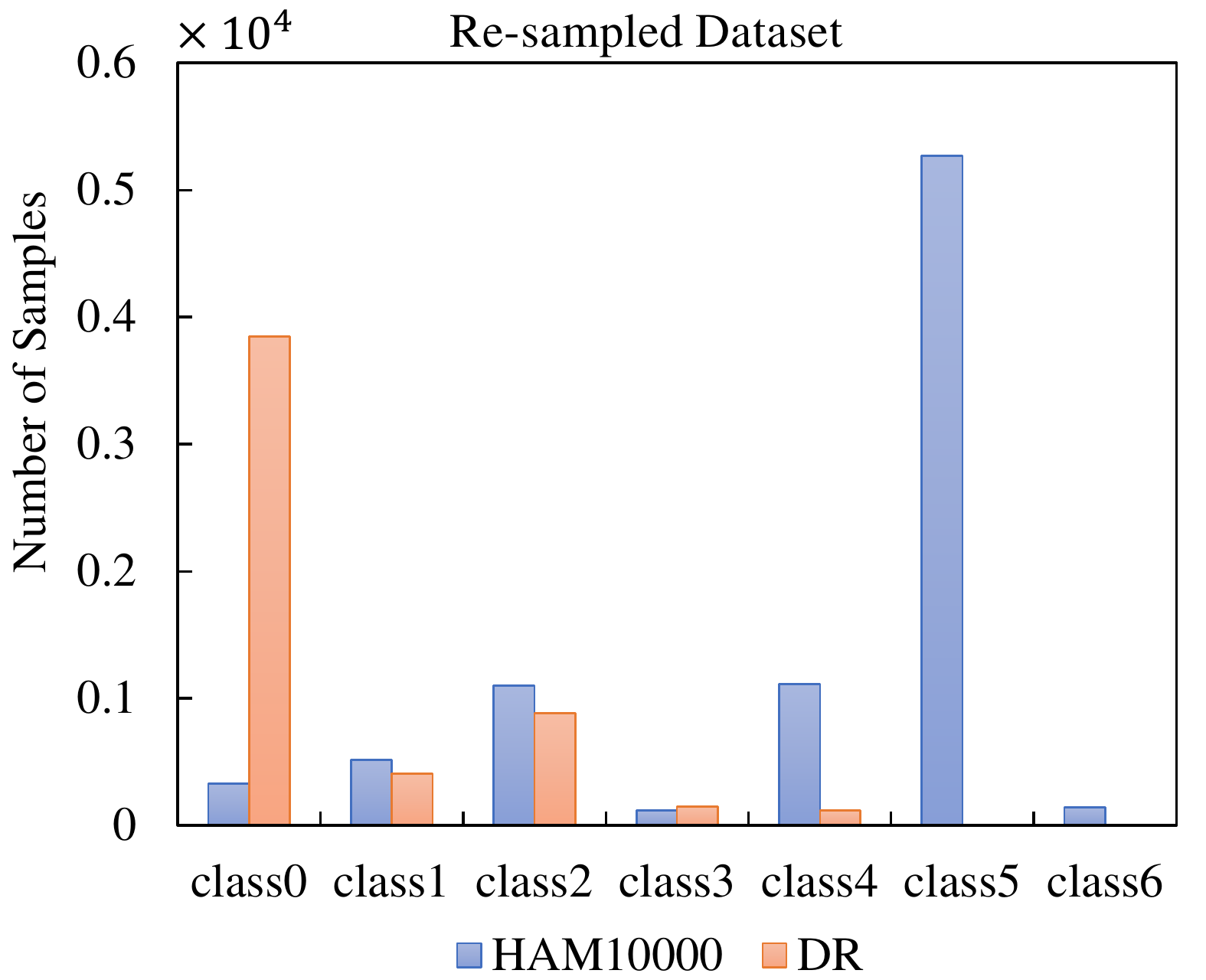}
	\label{fig:resampled_dataset}}
	\caption{The distribution of the two medical datasets considered as the target-domain dataset.}
\end{figure*}

\textit{\imagenet}\footnote{\url{https://www.image-net.org/}} is one of the largest publicly-available computer-vision datasets. It contains images belonging to $1,000$ categories subdivided in around $1.2$ million training images, $50,000$ validation images, and $100,000$ test images. The images are collected from the Internet by search engines and labeled by humans via crowdsourcing. We use this dataset as our source-domain dataset. We use its training set as our training dataset (we use models pre-trained on it) and its validation dataset to create our similarity encoder. To this end, we subdivide the validation dataset into $40,000$ samples that we use to train the similarity encoder and $10,000$ that we use to test its performances.\newline

\noindent \textit{\dr} (DR) \footnote{\url{https://www.kaggle.com/c/diabetic-retinopathy-detection/data}} is a medical dataset consisting of $35,126$ images with dimensions $4,652 \times 3,168$ and a label that ranges from 0 to 4, corresponding to the rating of the presence of diabetic retinopathy. We resize these samples to be $200 \times 200$. We perform re-sampling on the data samples to get a training/testing set of $3,000$/$2,400$ samples. The original and the re-sampled DR dataset distribution is represented as orange bars in the left and right plots of  Fig.~\ref{fig:resampled_dataset}.\newline

\noindent \textit{\md}\footnote{\url{https://dataverse.harvard.edu/dataset.xhtml?persistentId=doi:10.7910/DVN/DBW86T}} is a large collection of multi-source dermatoscopic images of common pigmented skin lesions, which includes $10,015$ samples of 7 types (\textit{``akiec'', ``bcc'', ``bkl'', ``df'', ``mel'', ``nv'', ``vasc''}) of skin cancer. The average image size is $600 \times 450$ pixels. We resize these data samples to be $200 \times 200$ pixels. Similarly to the  DR dataset, we perform re-sampling on this dataset. The collected training set contains $7,800$ samples, the testing set contains $780$ samples, and the distribution of re-sampled \md is represented as blue bar in the histograms in Fig.~\ref{fig:resampled_dataset}.\newline

\myparagraph{Preprocessing.} We rescale the input images in~$\mathcal{X} = [0, 1]^{d^{\prime}\times d^{\prime} \times 3}$ to match the input size $d$ of the considered models. In the process of generating adversarial queries, this requires padding input images with zeros.

\myparagraph{Classifiers.} We consider three different architectures pretrained on \imagenet and implemented on TensorFlow-Slim\footnote{\url{https://github.com/tensorflow/models/tree/master/research/slim\#Pretrained}} as target models: AlexNet~\cite{alex12-nips}, ResNet50~\cite{resnet50} and Inception-V3~\cite{inception-v3}. 
AlexNet\footnote{\url{https://drive.google.com/file/d/1ICnwX2fgyPMkJ0DyjOdLDadEO0C9C_ll/view}} and ResNet50\footnote{\url{http://download.tensorflow.org/models/resnet_v2_50_2017_04_14.tar.gz}} have input size $d = 224$, Inception-V3\footnote{\url{http://download.tensorflow.org/models/inception_v3_2016_08_28.tar.gz}} has input size $d = 299$. 

\myparagraph{Similarity Encoder.}
Following Chen et al.~\cite{chen2020stateful}, our similarity encoder is based on a three-layer CNN. The architecture of this CNN is represented in Table~\ref{tab:cnn-architecture}. In our experiments, we set the dimension of the space in which the samples are projected to $256$. 
To train the similarity encoder, as shown in Fig.~\ref{fig:encoder} and explained in details in Section \ref{sec:stateful-defenses}, we first create $p$ sample pairs from the validation dataset of \imagenet. Then, we train the encoder from scratch on these sample pairs using the RMSprop optimizer with a batch size of $32$, $100$ epochs, a learning rate of $10^{-4}$, and weight decay of $10^{-6}$. We set the $z$ parameter of Eq.~\eqref{eq:ls_encoder} to $1$. We are considering, as target models, classifiers with two different input dimensions ($224$ and $299$). Therefore, we trained two different similarity encoders to use a similarity encoder with the same input dimension as the considered classifier. We show their performance in Table.~\ref{tab:pretrained-encoder}.

\begin{table}[!t]
	\centering
	\caption{The architecture of the similarity encoder~\cite{chen2020stateful}.}
	\label{tab:cnn-architecture}
     \resizebox{0.40\textwidth}{!}
   {
	\begin{tabular}{l|c}
		\toprule
		Layer Type & Dimension\\
		\midrule
		Conv. + ReLU & 32 filters ($3\times3$) \\
		Conv. + ReLU & 32 filters ($3\times3$) \\
		Max Pooling & $2\times2$ \\
		Dropout & $p=0.25$ \\
		Conv. + ReLU & 64 filters ($3\times3$) \\
		Conv. + ReLU & 64 filters ($3\times3$) \\
		Max Pooling & $2\times2$ \\
		Dropout & $p=0.25$ \\
		Dense + ReLU  & 512 \\
		Dropout & $p=0.5$ \\
		Dense  & 256 \\
		\bottomrule
	\end{tabular}
	}
\end{table}

\begin{table}[!t]
	\centering
	\caption{The performance of the similarity encoder.}
	\label{tab:pretrained-encoder}
      \resizebox{0.40\textwidth}{!}
   {
	\begin{tabular}{c|c}
		\hline
		$d$ & Accuracy of Similarity Encoder \\ \hline
		224 &  62.60\%\\
		299 &  63.98\%\\ \hline
	\end{tabular}
 }
\end{table}

\myparagraph{Stateful Detection.}
Given that we choose the \imagenet dataset as our source domain, we use the \imagenet validation dataset to compute the threshold $\rho$ of the detector. Following~\cite{chen2020stateful}, we set $k=50$, and we employ the same procedure used by its authors to compute the detection threshold so that only a low and thus reasonable number of benign queries is flagged. This procedure sets the threshold so that if the entire set (constituted by benign samples) were randomly streamed as queries, the false positive rate would be $0.1\%$.

\myparagraph{Adversarial Reprogramming.}
To optimize the adversarial program $\vct \delta$, we use Algorithm~\ref{alg:reprogramming}. 
Before optimizing it, for the target-domain datasets, we fix $h$ as a MLM function that maps every 6 labels of the source dataset to one label of the target dataset, as explained in Sec. \ref{sec:adversarial-reprogramming}. We set the step size for updating the adversarial program parameters (\vct W) $\eta$ to $0.05$, and we use $N=10$ epochs. 
We consider DR, and \md as target-domain datasets, and we employ a batch size of  $B=24$ and $B=39$ for DR and \md samples, sampled randomly from the training set of the target-domain dataset $\mathcal{T}$.
To optimize the adversarial program in the white-box scenario, we set the learning rate $lr=0.05$. 
For the black-box scenario, as in \cite{tsai20-pmlr},
we set $b = d \times d \times 3$ and $\mu = 0.1$. In our experiments, we consider many different values for the parameter $q$.

\subsection{Experimental Results}
\label{sec:experiments_results}
In the following, we report the experimental results to assess the effectiveness of stateful defenses against adversarial reprogramming. We denote with R the accuracy obtained executing adversarial reprogramming in a white-box scenario and with BR the accuracy obtained in a black-box scenario. 

\myparagraph{White vs Black-box Reprogramming.} First, we compare the success rates of reprogramming programs generated in the black-box scenario with those generated in the white-box scenario. To this end, we employ three different models (AlexNet, ResNet50, and Inception-V3) and two medical datasets (DR and \md), fixing the parameter $q$ of DR and \md respectively to $q=55$ and $q=65$. We denote with $Tr$ ($Ts$) the dataset of the target domains we use to compute (test) the adversarial programs and the accuracy obtained reprogramming the target model in the white-box (black-box) scenario with $R_t$ ($BR_t$). As the sample for computing the adversarial program are usually difficult to collect for the attacker, in this experiment, we assess the performance for different numbers of training samples $Tr$. We present the result in Table~\ref{tab:black-white-performance}. 
From the Table~\ref{tab:black-white-performance}, we can see that $R_t$ is always greater than $BR_t$, and the difference, $R_t-BR_t$, between $R_t$ and $BR_t$ is located in $[0.33\%, 5.75\%]$. Moreover, we also notice that $R_t-BR_t$ is relatively large when the number of training samples is small. If we ignore the cases with a small number of training samples (\textit{pink values in Table~\ref{tab:black-white-performance}}), we obtain $R_t-BR_t\in [0.33\%, 1.81\%]$. Therefore, we can conclude that the performance of reprogramming queries generated in the black-box scenario is almost the same as the one generated in the white-box scenario when the training dataset is sufficiently large. 

\begin{table*}[!t]
	\centering
	\caption{Results of reprogramming AlexNet, ResNet50 and Inception-V3 for different sizes of the dataset employed by the attacker to compute the adversarial program ($Tr$). For each reprogramming task, the table reports the reprogramming accuracy obtained in a white box scenario ($R_t$), the reprogramming accuracy obtained in a black-box scenario ($BR_t$) and the difference between them ($ R_t-BR_t $).}
	\label{tab:black-white-performance}
   \resizebox{\textwidth}{!}
   {
	\begin{tabular}{c|cc|ccccccccc}
		\hline
		\multirow{3}{*}{Dataset} & \multirow{3}{*}{$Tr$} & \multirow{3}{*}{$Ts$} & \multicolumn{3}{c}{AlexNet} & \multicolumn{3}{c}{ResNet50} & \multicolumn{3}{c}{Inception-V3} \\ 
		&      &      & $R_t$   & $BR_t$  & $ R_t-BR_t $ & $R_t$   & $BR_t$  & $ R_t-BR_t $ & $R_t$   & $BR_t$ & $ R_t-BR_t $  \\ \hline
		\multirow{3}{*}{DR} & 6000 & 2400 & 80.70\% & 80.09\% &0.61\% & 80.16\% & 79.51\% &0.65\% & 80.13\% & 79.80\% &0.33\% \\
		& 3000 & 2400 & 80.33\% & 79.25\% &1.08\% & 79.36\% & 78.17\% &1.19\% & 79.84\% & 79.27\% &0.57\% \\
		& 1500 & 2400 & 79.81\% & 78.39\% &1.42\% & {\color[HTML]{EE5AA9}78.06\%} & {\color[HTML]{EE5AA9}74.78\%} &
		{\color[HTML]{EE5AA9}3.28\%} &  {\color[HTML]{EE5AA9}79.40\%} & {\color[HTML]{EE5AA9}76.68\%} &
		{\color[HTML]{EE5AA9}2.72\%} \\ \hline
		\multirow{3}{*}{\md} & 9200 & 780  & 81.07\% & 80.71\% &0.36\% & 81.55\% & 80.20\% &1.35\% & 80.32\% & 79.98\% &0.34\% \\
		& 7800 & 780  & 80.78\% & 79.72\% &1.06\% & 80.92\% & 79.11\% &1.81\% & 78.68\% & 77.52\% &1.16\% \\
		& 3900 & 780  & 77.24\% & 76.28\% &0.96\% & {\color[HTML]{EE5AA9}75.34\%} & {\color[HTML]{EE5AA9}71.77\%} &
		{\color[HTML]{EE5AA9}3.57\%} & {\color[HTML]{EE5AA9}74.42\%} & {\color[HTML]{EE5AA9}68.67\%} &
		{\color[HTML]{EE5AA9}5.75\%} \\ \hline
	\end{tabular}
 }
\end{table*}

\myparagraph{Defending against Black-box Reprogramming.} 
After, we assess the performance of the proposed detector against a black-box attack. In this experiment, we also evaluate the effect of tuning the parameter $q$ that influences the number of queries the attack will make to the target model. In this experiment, we consider AlexNet and ResNet50 as the target models, and we fix the number of training samples of DR and \md as $Tr=3,000$ and $Tr=7,800$, the number of testing dataset of DR and \md as $Ts=2,400$ and $Ts=780$. Then, we compute the program with different values of $q$ and present the result in Table~\ref{tab:detecting_performance1}. From this Table, we can see how the attack performance of adversarial reprogramming queries generated based on the target model ($BR_t$) positively correlates with the parameter $q$ and thus the number of queries. 

\myparagraph{Defending against fine-tuned Programs.} Finally, we test the effectiveness of our detector when the attacker first computes the adversarial program on a surrogate model and then refines it, making few queries to their target model. In this experiment, we consider AlexNet and ResNet50 as target and surrogate models. We set the number of training (testing) samples of DR and \md as in our previous experiment and vary $q$.
We denote the accuracy of the surrogate model in the white-box scenario with $R_s$.
We present the results in Table~\ref{tab:detecting_performance}. In this Table, we can see that when the attacker employs a surrogate model can obtain $BR_t \in [77.64\%, 82.21\%]$ and $BR_t \in [75.86\%, 81.19\%]$ for DR and \md when the surrogate model is applied. By comparing Table~\ref{tab:detecting_performance} with Table~\ref{tab:detecting_performance1}, we can see that by using a surrogate model, the attacker obtains a similar reprogramming accuracy, greatly reducing the number of queries $Q$ that it has to issue to the model. Moreover, we can see that when the attacker makes only few queries to the target model, the detector's performances (in light blue in the Table) are lower. Overall, we can conclude that, in this scenario, the considered statefull defense remains effective, although its effectiveness is reduced.

\begin{table*}[!t]
	\centering
	\caption{The reprogramming accuracy obtained reprogramming the target model in a black-box scenario ($BR_t$) for different values of the attack's hyperparameter $q$ that controls the number of queries $Q$ made by the attacker to the target model, and the corresponding performance of the detector, measured as the number of detection ($D$) and the percentage of detected attacks ($\sigma^*$) with respect to the number of possible detections.}
	\label{tab:detecting_performance1}
    \resizebox{0.75\textwidth}{!}
   {
	\begin{tabular}{c|c|ccccc}
	\hline
	Dataset & \begin{tabular}[c]{@{}c@{}}Target\\ Model\end{tabular} & $q$ & $BR_t$ & $Q$ & $D$ & $\sigma^{\star}$ \\ \hline
	\multirow{4}{*}{DR} & \multirow{2}{*}{AlexNet} & 45 & 77.54\% & 110400 & 1810 & 83.61\% \\
	 &  & 55 & 79.25\% & 134400 & 2266 & 85.99\% \\ \cline{2-7} 
	 & \multirow{2}{*}{ResNet50} & 45 & 76.56\% & 110400 & 1794 & 82.88\% \\
	 &  & 55 & 78.17\% & 134400 & 2237 & 84.89\% \\ \hline
	\multirow{4}{*}{\md} & \multirow{2}{*}{AlexNet} & 55 & 79.36\% & 43680 & 805 & 93.99\% \\
	 &  & 65 & 79.72\% & 51480 & 980 & 97.09\% \\ \cline{2-7} 
	 & \multirow{2}{*}{ResNet50} & 55 & 77.54\% & 43680 & 779 & 90.95\% \\
	 &  & 65 & 79.11\% & 51480 & 973 & 96.39\% \\ \hline
	\end{tabular}
 }
\end{table*}

\vspace{2.5em}

\begin{table*}[!t]
	\centering
	\caption{The performance of the adversarial programs computed on a surrogate model and fine-tuned with a few queries to the target model. We denote with $R_s$ the reprogramming accuracy obtained reprogramming the surrogate model in a white-box scenario, with $BR_t$ the reprogramming accuracy obtained fine-tuning the program by querying the target model with the black-box adversarial reprogramming attack, with $q$ the attack's hyperparameter that controls the number of queries $Q$, with $D$ the number of detection and with $\sigma^*$ the percentage of detected attacks with respect to the number of possible detections.}
	\label{tab:detecting_performance}
    \resizebox{\textwidth}{!}
   {
	\begin{tabular}{c|ccc|ccccc}
		\hline
		Dataset & \begin{tabular}[c]{@{}c@{}}Target\\ Model\end{tabular} & \begin{tabular}[c]{@{}c@{}}Surrogate\\ Model\end{tabular} & $R_s$ & $q$ & $BR_t$ & $Q$ & $D$ & $\sigma^{\star}$ \\ \hline
		\multirow{4}{*}{DR} 
		& \multirow{2}{*}{AlexNet} & \multirow{2}{*}{ResNet50} & \multirow{2}{*}{79.36\%} & 5 & 79.96\% & 14400 & 208 & {\color[HTML]{6665CD}73.67\%} \\
		&  &  &  & 10 & 80.29\% & 26400 & 452 & 87.32\% \\ \cline{2-9} 
		& \multirow{2}{*}{ResNet50} & \multirow{2}{*}{AlexNet} & \multirow{2}{*}{80.33\%} & 5 & 77.64\% & 14400 & 193 & {\color[HTML]{6665CD}68.35\%} \\
		&  &  &  & 10 & 82.21\% & 26400 & 461 & 89.06\% \\ \hline
		\multicolumn{1}{l|}{\multirow{4}{*}{\md}} 
		& \multirow{2}{*}{AlexNet} & \multirow{2}{*}{ResNet50} & \multirow{2}{*}{80.92\%} & 5 & 78.71\% & 4680 & 78 & {\color[HTML]{6665CD}85.00\%} \\
		\multicolumn{1}{l|}{} &  &  &  & 10 & 81.19\% & 8580 & 148 & 87.97\% \\ \cline{2-9} 
		\multicolumn{1}{l|}{} & \multirow{2}{*}{ResNet50} & \multirow{2}{*}{AlexNet} & \multirow{2}{*}{80.78\%} & 5 & 75.86\% & 4680 & 73 & {\color[HTML]{6665CD}79.55\%} \\
		\multicolumn{1}{l|}{} &  &  &  & 10 & 80.43\% & 8580 & 144 & 85.59\% \\ \hline
	\end{tabular}
 }
\end{table*}

\section{Related Work}
\label{sec:related_work}
In this section, we briefly review related work on adversarial reprogramming. We then focus on the defense of the adversarial attacks in the black-box scenario.

\subsection{Adversarial Reprogramming} 
Adversarial reprogramming has been originally proposed in~\cite{elsayed19-ICLR}.
The authors have empirically assessed the performance of adversarial reprogramming using different trained and untrained deep neural networks. They showed that reprogramming usually fails when applied to untrained networks (\ie, neural networks with random weights), whereas it works when the target model is trained. In the latter case, reprogramming works even when the attacker can manipulate only a small subset of the image pixels. In~\cite{zheng2021adversarial} the authors have developed a first-order linear model of adversarial reprogramming to analyze the factors that affect its success. They show that reprogramming can fail and that its success inherently depends on the size of the average input gradient, which grows when input gradients for the target model are more aligned, and inputs have higher dimensionality. The authors of~\cite{tsai20-pmlr} have shown that adversarial reprogramming also works in black-box scenarios where the attacker has a query only access to its target model. Moreover, they have demonstrated reprogramming can be particularly beneficial in tasks with scarce data, as in that case, it can achieve even better performance than fine-tuning. No defense has been proposed against reprogramming in a black box scenario; therefore, in this work, we assess to which extent a defense previously proposed against evasion attacks can defend against adversarial reprogramming. 

\subsection{Previously Proposed Defenses} 
To the best of our knowledge, only one defense \cite{wang2019protecting} has been proposed against adversarial reprogramming. This defense, named hierarchical random switching (HRS), randomizes the model at test time to prevent adversaries from exploiting fixed model structures and parameters for malicious purposes. However, this defense has been tested only against white-box adversarial reprogramming attacks, that is not a realistic scenario. 
Defenses have been proposed against evasion attacks for the black-box scenarios: an attack that computes a perturbation that, if applied to a single sample, allows the attacker to have it classified as the desired class. It is worth noting that evasion attacks are less challenging than adversarial reprogramming because the attacker computes a perturbation ad-hoc for each input image. Differently, in reprogramming attacks, a single perturbation should allow having all the test images classified as belonging to a target class. 
Most defenses against black-box evasion attacks examine each query singularly (stateless detection), usually by checking if this query lays out the distribution of normal/benign data \cite{feinman17-arxiv, grosse_statistical_2017, metzen_detecting_2017}. 
However, effective detection under this stateless threat model has proven to be difficult~\cite{carlini17-aisec}. For this reason, Chen et al.~\cite{chen2020stateful} developed a stateful defense that jointly considers all the queries received by the classifier. In this work, we have shown at which extent this defense is effective against adversarial reprogramming in the black-box scenario.

\section{Contributions and Limitations of this work}
\label{sec:conclusion}

In this work, we addressed the problem of defending machine learning models against adversarial reprogramming in a black-box scenario. 

To the best of our knowledge, this is the first work proposing a countermeasure for this attack considering the realistic scenario where the target model is unknown to the attacker. Therefore, our stateful defence provides users for the first time a simple to tool to mitigate this threat.
We assessed the effectiveness of statefull defenses against this attack by experiments. To this end, a similarity encoder has been trained to map the adversarial queries to a low-dimensional space. In this space, we flag as adversarial the queries quite similar to other queries previously made by the same user.
Our experimental analysis shows that a large percentage of the queries made by the attacker to compute the adversarial program are flagged by our defense as adversarial. Once a single query has been detected as malicious, the account of the attacker can be blocked. Therefore, the attacker will have to create many different accounts to perpetrate the attack. Stateful defenses are thus highly effective for increasing the attacker's cost in this back-box scenario and, consequently, they represent a good deterrence mechanism. 
Even if the attacker exploits the transferability property of the attacks to reduce the number of detected queries and thus her effort, our experiments show that the proposed defence is still effective.

The main limitation of our work is that we have not tested the effectiveness of the statefull detector against an adaptive attacker who is aware of our defence and tries to evade it. Moreover, as in \cite{chen2020stateful} we have considered an infinite memory buffer, whereas, in practice, the defender should set the memory buffer lenght according to the system capacity. Analyzing the effectiveness of our defence in these scenarios is one of our future research directions.

\section*{Acknowledgments}
This work was partly supported by the PRIN 2017 project RexLearn, funded by the Italian Ministry of Education, University and Research (grant no. 2017TWNMH2); by BMK, BMDW, and the Province of Upper Austria in the frame of the COMET Programme managed by FFG in the COMET Module S3AI; by the Horizon 2020 project Starlight, founded by the European Union; and by the Key Research and Development Program of Shaanxi (Program Nos. 2021ZDLGY15-01, 2021ZDLGY09-04, 2021GY-004 and 2020GY-050), the International Science and Technology Cooperation Research Project of Shenzhen (GJHZ20200731095204013), the National Natural Science Foundation of China (Grant No. 61772419).

\bibliography{addref,bibDB}

\end{document}